\documentclass[aps,preprint]{revtex4}%
\usepackage{amsfonts}
\usepackage{amsmath}
\usepackage{amssymb}
\usepackage{graphicx}%
\providecommand{\U}[1]{\protect\rule{.1in}{.1in}}

\begin{document}
\title{Observational constraints on holographic tachyonic dark energy in interaction
with dark matter}
\author{Sandro M. R. Micheletti \footnote{smrm@fma.if.usp.br }}
\affiliation{Instituto de F\'{\i}sica, Universidade de S\~{a}o Paulo, CP 66318, 05315-970,
Sao Paulo, Brazil}

\begin{abstract}
We discuss an interacting tachyonic dark energy model in the context of the
holographic principle. The potential of the holographic tachyon field in
interaction with dark matter is constructed. The model results are compared
with CMB shift parameter, baryonic acoustic oscilations, lookback time and the
Constitution supernovae sample. The coupling constant of the model is
compatible with zero, but dark energy is not given by a cosmological constant.

\end{abstract}
\maketitle


\thispagestyle{empty}

In the last years, there have been several papers where an interaction in the
dark sector of the universe is considered \cite{1} - \cite{sandro}. A
motivation to considering the interaction is that dark energy and dark matter
will evolve coupled to each other, alleviating the coincidence problem
\cite{1}. A further motivation is that, assuming dark energy to be a field, it
would be more natural that it couples with the remaining fields of the theory,
in particular with dark matter, as it is quite a general fact that different
fields generally couple. In other words, it is reasonable to assume that there
is no symmetry preventing such a coupling between dark energy and dark matter
fields. Using a combination of several observational datasets, as supernovae
data, CMB shift parameter, BAO, etc., it has been found that the coupling
constant is small but non vanishing within at least $1\sigma$ confidence level
\cite{1}, \cite{3}, \cite{sandro}, \cite{sandro2}. In two recent works, the
effect of an interaction between dark energy and dark matter on the dynamics
of galaxy clusters was investigated through the Layser-Irvine equation, the
relativistic equivalent of virial theorem \cite{peebles}. Using galaxy cluster
data, it has been shown that a non vanishing interaction is preferred to
describe the data within several standard deviations \cite{virial}. However,
in most of these papers, the interaction term in the equation of motion is
derived from phenomenological arguments. It is interesting to obtain the
interaction term from a field theory. Some works have already taken a step in
such a direction \cite{sandro2}, \cite{amendola}. On the other hand, there
have been several papers where the dark energy is associated with the tachyon
scalar field. The tachyon field has been studied in recent years in the
context of string theory, as a low energy effective theory of D-branes and
open strings \cite{sen}. The pressure of the tachyon fluid is negative, and it
has been used in cosmology as a candidate to dark energy \cite{sandro2},
\cite{padmanabhan} - \cite{taqholo}. The first question about tachyons
concerns the choice of the potential. Common choices for the tachyonic
potential are the power law and the exponential potentials, both capable of
reproducing the recent period of accelerated expansion, the last of these
being motivated by some string theoretical models. However, these choices are
in fact arbitrary. In principle, any other form for the potential which leads
to recent accelerated expansion would be acceptable.

On the other hand, it is possible that a complete understanding of the nature
of dark energy will only be possible within a quantum gravity theory context.
Although results for quantum gravity are still missing, or at least premature,
it is possible to introduce, phenomenologically, some of its principles in a
model of dark energy. Recently, a combination of the tachyon model with the
holographic dark energy model has been made available \cite{taqholo} -
previously, combinations of quintessence and quintom models with holographic
dark energy had been proposed \cite{escalarholo}, \cite{quintomholo}.
Specifically, by imposing that the energy density of the tachyon fluid must
match the holographic dark energy density, namely $\rho_{\Lambda}=3c^{2}%
M_{Pl}^{2}L^{-2}$, where $c$ is a numerical constant and $L$ is the infrared
cutoff, it was demonstrated that the equation of motion of tachyons for the
non-interacting case reproduces the equation of motion for holographic dark
energy. In fact, to impose that the energy density of tachyons must match the
holographic dark energy density corresponds to specify the potential of
tachyons. This can be seen as a physical criterion to choose the potential.
Here, we generalize this idea for the interacting case.

We consider the lagrangian density%
\[
\mathcal{%
\mathcal{L}%
}_{M}\left(  x\right)  =\sqrt{-g}\left\{  -V(\varphi)\sqrt{1-\alpha
\partial^{\mu}\varphi\partial_{\mu}\varphi}+\frac{i}{2}[\bar{\Psi}\gamma^{\mu
}\nabla_{\mu}\Psi-\bar{\Psi}\overleftarrow{\nabla}_{\mu}\gamma^{\mu}%
\Psi]-(M-\beta\varphi)\bar{\Psi}\Psi\right\}  \text{ ,}%
\]
where $\alpha$ is a constant with dimension $MeV^{-4}$, $\beta$ is the
dimensionless coupling constant, $V(\varphi)$ the potential and $g$ the
determinant of the metric. From a variational principle, we obtain%
\begin{equation}
i\gamma^{\mu}\nabla_{\mu}\Psi-M^{\ast}\Psi=0\text{ ,} \label{dirac}%
\end{equation}%
\begin{equation}
i(\nabla_{\mu}\bar{\Psi})\gamma^{\mu}+M^{\ast}\bar{\Psi}=0\text{ ,}
\label{diracadj}%
\end{equation}
where $M^{\ast}\equiv M-\beta\varphi$, and%
\begin{equation}
\nabla_{\mu}\partial^{\mu}\varphi+\alpha\frac{\partial^{\mu}\varphi
(\nabla_{\mu}\partial_{\sigma}\varphi)\partial^{\sigma}\varphi}{1-\alpha
\partial_{\mu}\varphi\partial^{\mu}\varphi}+\frac{1}{\alpha}\frac
{dlnV(\varphi)}{d\varphi}=\frac{\beta\bar{\Psi}\Psi}{\alpha V(\varphi)}%
\sqrt{1-\alpha\partial^{\mu}\varphi\partial_{\mu}\varphi}\text{ .}
\label{eqmov_taquions}%
\end{equation}
The eq. (\ref{dirac}) and (\ref{diracadj}) are, respectively, the covariant
Dirac equation and its adjoint, in the case of a non vanishing interaction
between the Dirac field and the tachyon field $\varphi$. For homogeneous
fields and adopting the Friedmann-Robertson-Walker metric, $g_{\mu\nu}%
$=diag$\left(  1,-a^{2}\left(  t\right)  ,-a^{2}\left(  t\right)
,-a^{2}\left(  t\right)  \right)  $, where $a^{2}\left(  t\right)  $ is the
scale factor, the eq. (\ref{dirac}) and (\ref{diracadj}) lead to%
\begin{equation}
\frac{d(a^{3}\bar{\Psi}\Psi)}{dt}=0 \label{conser_psibarpsi}%
\end{equation}
and (\ref{eqmov_taquions}) reduces to%
\begin{equation}
\ddot{\varphi}=-(1-\alpha\dot{\varphi}^{2})\left[  \frac{1}{\alpha}%
\frac{dlnV(\varphi)}{d\varphi}+3H\dot{\varphi}-\frac{\beta\bar{\Psi}\Psi
}{\alpha V(\varphi)}\sqrt{1-\alpha\dot{\varphi}^{2}}\right]  \text{ ,}
\label{homotaq}%
\end{equation}
where $H\equiv\frac{\dot{a}}{a}$ is the Hubble parameter. From
(\ref{conser_psibarpsi}) we have\textbf{ }$\bar{\Psi}\Psi=\bar{\Psi}_{0}%
\Psi_{0}\left(  \frac{a_{0}}{a}\right)  ^{3}$.

From the energy-momentum tensor, we get%
\begin{align}
\rho_{\varphi}  &  =\frac{V(\varphi)}{\sqrt{1-\alpha\dot{\varphi}^{2}}}\text{
,}\label{rofi}\\
P_{\varphi}  &  =-V(\varphi)\sqrt{1-\alpha\dot{\varphi}^{2}}\text{
,}\label{pfi}\\
\rho_{\Psi}  &  =M^{\ast}\bar{\Psi}\Psi\text{ ,}\nonumber\\
P_{\Psi}  &  =0\text{ .}\nonumber
\end{align}
Notice that from (\ref{pfi}) and (\ref{rofi}) we have $\omega_{\varphi}%
\equiv\frac{P_{\varphi}}{\rho_{\varphi}}=\alpha\dot{\varphi}^{2}-1$. Deriving
(\ref{rofi}) and (\ref{pfi}) with respect to time and using (\ref{homotaq})
and (\ref{conser_psibarpsi}), we get%
\begin{equation}
\dot{\rho}_{\varphi}+3H\rho_{\varphi}(\omega_{\varphi}+1)=\beta\dot{\varphi
}\bar{\Psi}_{0}\Psi_{0}\left(  \frac{a_{0}}{a}\right)  ^{3}
\label{conser_rofi}%
\end{equation}
and%
\begin{equation}
\dot{\rho}_{\Psi}+3H\rho_{\Psi}=-\beta\dot{\varphi}\bar{\Psi}_{0}\Psi
_{0}\left(  \frac{a_{0}}{a}\right)  ^{3}\text{ ,} \label{conser_ropsi}%
\end{equation}
where the dot represents derivative with respect to time. The Friedmann
equation for a flat universe reads%
\begin{equation}
H^{2}=\frac{1}{3M_{Pl}^{2}}\left[  M^{\ast}\bar{\Psi}_{0}\Psi_{0}\left(
\frac{a_{0}}{a}\right)  ^{3}+\frac{V(\varphi)}{\sqrt{1-\alpha\dot{\varphi}%
^{2}}}\right]  \text{ ,} \label{friedmann}%
\end{equation}
where $M_{Pl}\equiv\left(  8\pi G\right)  ^{-1/2}$ is the reduced Planck mass.

In order to determine the dynamics of the interacting tachyon, it is necessary
to specify the potential $V(\varphi)$. In \cite{sandro2}, a power law
potential had been chosen. Here, instead of choosing an explicit form for
$V(\varphi)$, we will specify it implicitly, by imposing that the energy
density of the tachyon fluid, given by (\ref{rofi}), must match the
holographic dark energy density, $\rho_{\Lambda}=3c^{2}M_{Pl}^{2}L^{-2}$,
where $c$ is a numerical constant and $L$ is the infrared cutoff. The
evolution of the interacting tachyon fluid with redshift will be given by the
equation of evolution for the holographic dark energy density, with a certain
expression for the equation of state parameter $\omega_{\varphi}$. In fact, we
will see that imposing the energy density of tachyons to match the holographic
dark energy density leads to an expression for the potential of tachyons.

In \cite{Li} it has been argued that, in order that holographic dark energy
drives the recent period of accelerated expansion, the IR cutoff $L$ must be
the event horizon $R_{h}$. Substituting $R_{h}$ in the expression of the
holographic dark energy, we get $R_{h}=\frac{c}{H\sqrt{\Omega_{\varphi}}}$,
therefore,%
\[
\int_{t}^{\infty}\frac{dt^{\prime}}{a\left(  t^{\prime}\right)  }=\frac
{c}{a\left(  t\right)  H\sqrt{\Omega_{\varphi}}}\text{ .}%
\]
Differentiating both sides with respect to time, using the Friedmann equation
(\ref{friedmann}) together with conservation equations (\ref{conser_rofi}) and
(\ref{conser_ropsi}), we obtain%
\begin{equation}
\frac{d\Omega_{\varphi}}{dz}=-\frac{2\Omega_{\varphi}}{1+z}\left[  \frac
{\sqrt{\Omega_{\varphi}}}{c}+\frac{3\Omega_{\varphi}\omega_{\varphi}+1}%
{2}\right]  \text{ .} \label{eq_mov_holo}%
\end{equation}
Equation (\ref{eq_mov_holo}) is just the equation of evolution for the
holographic dark energy \cite{Li}.

We define $r\equiv\frac{\rho_{\Psi}}{\rho_{\varphi}}$. Deriving $r$ with
respect to time, using (\ref{conser_rofi}), (\ref{conser_ropsi}) and
eliminating $\dot{\varphi}$ by using $\dot{\varphi}=\pm\sqrt{\frac
{1+\omega_{\varphi}}{\alpha}}$, we obtain%
\begin{equation}
\dot{r}=3Hr\omega_{\varphi}-sign\left[  \dot{\varphi}\right]  \frac
{\beta\left(  1+r\right)  ^{2}\sqrt{1+\omega_{\varphi}}}{3M_{Pl}^{2}%
\sqrt{\alpha}H^{2}}\bar{\Psi}_{0}\Psi_{0}\left(  \frac{1+z}{1+z_{0}}\right)
^{3}\text{ .} \label{rdot}%
\end{equation}
The sign of $\dot{\varphi}$ is arbitrary, as it can be modified by
redefinitions of the field, $\varphi\rightarrow-\varphi$, and of the coupling
constant, $\beta\rightarrow-\beta$. We can rewrite $\bar{\Psi}_{0}\Psi_{0}$ in
terms of observable quantities. In fact, by imposing that the dark matter
density today matches the observed value, we obtain $M\bar{\Psi}_{0}\Psi
_{0}=\frac{3M_{Pl}^{2}H_{0}^{2}\left(  1-\Omega_{\phi0}\right)  }%
{1-\frac{\beta}{M\sqrt{\alpha}}\phi_{0}}$, where we defined $\phi\equiv
\sqrt{\alpha}\varphi$. Furthermore, noticing that $r=\frac{1-\Omega_{\varphi}%
}{\Omega_{\varphi}}$, we can eliminate $r$ and $\dot{r}$ in favor of
$\Omega_{\varphi}$ and $\dot{\Omega}_{\varphi}$ in (\ref{rdot}). Using
(\ref{eq_mov_holo}) we obtain, after some algebra%
\begin{equation}
\omega_{\phi}\left(  z\right)  =-\frac{1}{3}-\frac{2\sqrt{\Omega_{\phi}\left(
z\right)  }}{3c}+\frac{\gamma\left(  z\right)  }{18}\left[  \gamma\left(
z\right)  +\sqrt{\gamma\left(  z\right)  ^{2}+24\left(  1-\frac{\sqrt
{\Omega_{\phi}\left(  z\right)  }}{c}\right)  }\right]  \text{ ,}
\label{wfi_final}%
\end{equation}
where%
\[
\gamma\left(  z\right)  \equiv\delta\frac{1-\Omega_{\phi0}}{H_{0}\Omega_{\phi
}\left(  z\right)  E^{3}\left(  z\right)  }\left(  \frac{1+z}{1+z_{0}}\right)
^{3}\text{ ,}%
\]
with%
\begin{equation}
E\left(  z\right)  \equiv\frac{H\left(  z\right)  }{H_{0}}=\sqrt{\left[
1-\delta\Delta\phi\left(  z\right)  \right]  \frac{1-\Omega_{\phi0}}%
{1-\Omega_{\phi}\left(  z\right)  }\left(  \frac{1+z}{1+z_{0}}\right)  ^{3}%
}\text{ ,} \label{E_final}%
\end{equation}
where $\Delta\phi\left(  z\right)  \equiv\phi\left(  z\right)  -\phi_{0}$ and
$\delta\equiv\frac{\frac{\beta}{M\sqrt{\alpha}}}{1-\frac{\beta}{M\sqrt{\alpha
}}\phi_{0}}$ is an effective coupling constant. Notice that, if $\delta=0$,
(\ref{wfi_final}) reproduces the equation of state parameter obtained in
\cite{Li}.

The evolution of the tachyon scalar field is given by%
\begin{equation}
\frac{d\phi}{dz}=-\frac{\sqrt{1+\omega_{\phi}\left(  z\right)  }}%
{H_{0}E\left(  z\right)  \left(  1+z\right)  }\text{ .} \label{eq_fi}%
\end{equation}
From (\ref{eq_mov_holo}) and (\ref{eq_fi}) we can calculate the evolution with
redshift of all observables in the model. If we wish to calculate the time
dependence, we need to integrate the Friedmann equation (\ref{friedmann}),
which can be written in the form%
\[
\frac{dt}{dz}=-\frac{1}{H_{0}E(z)(1+z)}\text{ .}%
\]

Here it is worth saying that in the holographic dark energy model, in the non
interacting case - (\ref{wfi_final}) with $\delta=0$ - $\omega_{\phi}$ can be
less than $-1$. However, as already mentioned in \cite{taqholo}, if we wish
that the holographic dark energy is the tachyon, then because (\ref{eq_fi}),
$\omega_{\phi}$ must be more than $-1$. Nevertheless, in the interacting case
considered here, due to the fact that $\omega_{\phi}$ depends explicitly on
$\phi$, $\omega_{\phi}$ can not be less than $-1$. On the other hand, the
square root in (\ref{wfi_final}) must be real. We can verify that
$\omega_{\phi}$ is real and $\omega_{\phi}>-1$ if (i) $\frac{\sqrt
{\Omega_{\phi0}}}{c}<1$ or (ii) $\frac{\sqrt{\Omega_{\phi0}}}{c}>1$ and
$\frac{\left\vert \delta\right\vert }{H_{0}}>2\frac{\Omega_{\phi0}}%
{1-\Omega_{\phi0}}\sqrt{6\left(  \frac{\sqrt{\Omega_{\phi0}}}{c}-1\right)  }$.
However, case (ii) is irrelevant, as it corresponds to large values of
$\frac{\left\vert \delta\right\vert }{H_{0}}$. For example, if $\Omega_{\phi
0}=0.7$ and $c=0.8$, we have $\frac{\left\vert \delta\right\vert }{H_{0}%
}\gtrsim2.45$. Below, we will see that the observational data constrain
$\frac{\left\vert \delta\right\vert }{H_{0}}\sim10^{-1}$. In order that
$\omega_{\phi}$ be real for all future times, as $\Omega_{\phi}\rightarrow1$,
it is necessary that $c\geq1$.

It is interesting to notice that the condition $\frac{\sqrt{\Omega_{\phi0}}%
}{c}<1$ is precisely the condition for which the entropy of the universe
increases \cite{Li}. As $\Omega_{\phi}\rightarrow1$ in the future, it is
necessary that $c\geq1$. Therefore, the condition for $\omega_{\phi}$ be real
is precisely the same one for the entropy to increase. So the model respects
the second law of thermodynamics.

\begin{figure}[pth]
\begin{center}
\includegraphics[width=8cm,height=6.5cm]{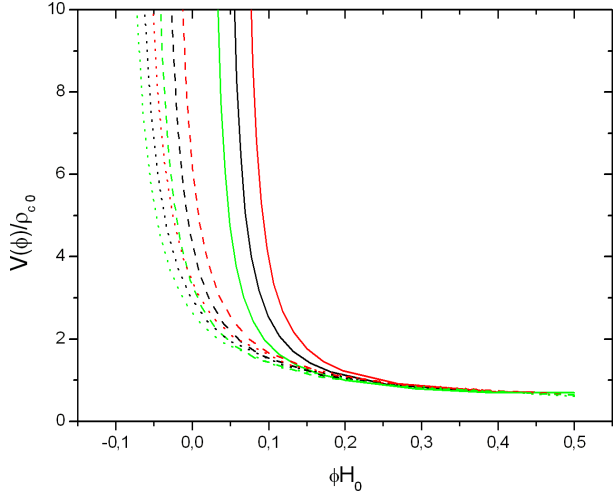}
\end{center}
\caption{Potential of tachyons $V(\phi)$, in units of $\rho_{c0}=3M_{Pl}%
^{2}H_{0}^{2}$. $\phi$ is in units of $H_{0}^{-1}$. The solid lines are for
$c=0.85$, the dashed ones are for $c=1.1$ and the dotted are for $c=1.35$. For
each value of $c$ the curves from right to left are for $\frac{\delta}{H_{0}%
}=-0.1$ (red), $\frac{\delta}{H_{0}}=0$ (black) and $\frac{\delta}{H_{0}%
}=+0.1$ (green), respectively.}%
\label{fig_potential}%
\end{figure}

From (\ref{rofi}), we can compute the potential $V\left(  z\right)  $ as%
\begin{equation}
\frac{V\left(  z\right)  }{\rho_{c0}}=E^{2}\left(  z\right)  \Omega_{\phi
}\left(  z\right)  \sqrt{-\omega_{\phi}\left(  z\right)  }\text{ ,}
\label{potential}%
\end{equation}
where $\rho_{c0}=3M_{Pl}^{2}H_{0}^{2}$, $E\left(  z\right)  $ is given by
(\ref{E_final}), $\omega_{\phi}\left(  z\right)  $ is given by
(\ref{wfi_final}) and $\Omega_{\phi}\left(  z\right)  $ is the solution of
(\ref{eq_mov_holo}). From (\ref{potential}) and (\ref{eq_fi}), we can compute
$V(\phi)$. In figure \ref{fig_potential}, $V(\phi)$ is shown for some values
of $\delta$ and $c$. Notice that, as we chose $\dot{\phi}$ positive, then
$\phi$ evolves to the mininum of the potential. However, if we had chosen
$\dot{\phi}$ negative, then because the right hand side of (\ref{eq_fi}) would
has the opposite sign, $V(\phi)$ would be now an increasing function of $\phi$
and again $\phi$ would evolve to the mininum of the
potential.

The equation for evolution of $\phi$ (\ref{eq_fi}) can be written in an
integral form as%
\[
\Delta\phi=-\frac{1}{H_{0}}\int_{0}^{z}\frac{\sqrt{1+\omega_{\phi}\left(
z\right)  }}{E(z)(1+z)}dz\text{ .}%
\]
Since the model depends on $\Delta\phi$ - through $E\left(  z\right)  $ - and
neither on $\phi$ nor on $\phi_{0}$, then it is independent of $\phi_{0}$. In
other words, $\phi_{0}$ is not a parameter of the model and can be chosen
arbitrarily. Therefore, the parameters of the model are $\delta$, $c$, $h$ and
$\Omega_{\phi0}$. Below, we discuss the comparison with observational data and
the results obtained.

In \cite{lookback}, the lookback time method has been discussed. Given an
object $i$ at redshift $z_{i}$, its age $t(z_{i})$ is defined as the
difference between the age of the universe at $z_{i}$ and the age of the
universe at the formation redshift of the object, $z_{F}$, that is,
\begin{align}
t(z_{i})  &  =H_{0}^{-1}\left[  \int_{z_{i}}^{\infty}\frac{dz^{\prime}%
}{(1+z^{\prime})E(z^{\prime})}-\int_{z_{F}}^{\infty}\frac{dz^{\prime}%
}{(1+z^{\prime})E(z^{\prime})}\right] \nonumber\\
&  =H_{0}^{-1}\int_{z_{i}}^{z_{F}}\frac{dz^{\prime}}{(1+z^{\prime}%
)E(z^{\prime})}=t_{L}(z_{F})-t_{L}(z_{i})\text{ ,} \label{age}%
\end{align}
where $t_{L}$ is the lookback time, given by
\[
t_{L}(z)=H_{0}^{-1}\int_{0}^{z}\frac{dz^{\prime}}{(1+z^{\prime})E(z^{\prime}%
)}\text{ .}%
\]
Using (\ref{age}), the observational lookback time $t_{L}^{obs}(z_{i})$ is
\begin{align}
t_{L}^{obs}(z_{i})  &  =t_{L}(z_{F})-t(z_{i})=[t_{0}^{obs}-t(z_{i}%
)]-[t_{0}^{obs}-t_{L}(z_{F})]\nonumber\\
&  =t_{0}^{obs}-t(z_{i})-df\text{ ,} \label{lookobs}%
\end{align}
where $t_{0}^{obs}$ is the estimated age of the universe today and $df$ is the
delay factor,
\[
df\equiv t_{0}^{obs}-t_{L}(z_{F})\ .
\]
We now minimize $\chi_{lbt}^{2}$,
\[
\chi_{lbt}^{2}=\sum_{i=1}^{N}\frac{[t_{L}(z_{i},\vec{p})-t_{L}^{obs}%
(z_{i})]^{2}}{\sigma_{i}^{2}+\sigma_{t_{0}^{obs}}^{2}}\text{ ,}%
\]
where $t_{L}(z_{i},\vec{p})$ is the theoretical value of the lookback time in
$z_{i}$, $\vec{p}$ denotes the theoretical parameters, $t_{L}^{obs}(z_{i})$ is
the corresponding observational value given by (\ref{lookobs}), $\sigma_{i}$
is the uncertainty in the estimated age $t(z_{i})$ of the object at $z_{i}$,
which appears in (\ref{lookobs}) and $\sigma_{t_{0}^{obs}}$ is the uncertainty
in getting $t_{0}^{obs}$. The delay factor $df$ appears because of our
ignorance about the redshift formation $z_{F}$ of the object and has to be
adjusted. Note, however, that the theoretical lookback time does not depend on
this parameter, and we can marginalize over it.

In \cite{age35} and \cite{age32} the ages of 35 and 32 red galaxies are
respectively given. For the age of the universe one can adopt $t_{0}%
^{obs}=13.73\pm0.12Gyr$ \cite{wmap5yr}. Although this estimate for
$t_{0}^{obs}$ has been obtained assuming a $\Lambda CDM$ universe, it does not
introduce systematical errors in the calculation: any systematical error
eventually introduced here would be compensated by the adjust of $df$, in
(\ref{lookobs}). On the other hand, such an estimate is in perfect agreement
with other estimates, which are independent of the cosmological model, as for
example $t_{0}^{obs}=12.6_{-2.4}^{+3.4}Gyr$, obtained from globular cluster
ages \cite{krauss} and $t_{0}^{obs}=12.5\pm3.0Gyr$, obtained from
radioisotopes studies \cite{cayrel}.

For the cosmic radiation shift parameter in the flat universe we have
\[
R=\sqrt{\Omega_{M}}\int_{0}^{z_{ls}}\frac{dz^{\prime}}{E(z^{\prime})}\ ,
\]
where $z_{ls}$ is the last scattering surface redshift parameter. The value
$R$ has been estimated from the 5-years WMAP \cite{wmap5yr} results as
$R_{obs}=1.715\pm0.021$, for the flat universe, with $z_{ls}=1090.5\pm1.0$ and
is very weakly model dependent \cite{R}. Thus we add to $\chi^{2}$ the term%
\[
\chi_{CMB}^{2}=\frac{\left(  R-R_{obs}\right)  ^{2}}{\sigma_{R}^{2}}\ \text{.}%
\]

Baryonic Acoustic Oscilations (BAO) are described in terms of the parameter
\[
A=\sqrt{\Omega_{M}}E(z_{BAO})^{-1/3}\left[  \frac{1}{z_{BAO}}\int_{0}%
^{z_{BAO}}\frac{dz^{\prime}}{E(z^{\prime})}\right]  ^{2/3}\text{ ,}%
\]
where $z_{BAO}=0.35$. It has been estimated that $A_{obs}=0.493\pm0.017$
\cite{BAO1}. We thus add to $\chi^{2}$ the term
\[
\chi_{BAO}^{2}=\frac{\left(  A-A_{obs}\right)  ^{2}}{\sigma_{A}^{2}}\ \text{.}%
\]

The BAO distance ratio $r_{BAO}\equiv D_{V}\left(  z=0.35\right)
/D_{V}\left(  z=0.20\right)  =1.812\pm0.060$, estimated from the joint
analysis of the 2dFGRS and SDSS\ data \cite{BAO2}, has also been included. It
was demonstrated in \cite{BAO2} that this quantity is weakly model dependent.
The quantity $D_{V}\left(  z_{BAO}\right)  $ is given by%
\[
D_{V}\left(  z_{BAO}\right)  =c\left[  \frac{z_{BAO}}{H\left(  z_{BAO}\right)
}\left(  \int_{0}^{z_{BAO}}\frac{dz^{\prime}}{H\left(  z^{\prime}\right)
}\right)  ^{2}\right]  ^{1/3}\text{ .}%
\]
So we have the contribution%
\[
\chi_{r_{BAO}}^{2}=\frac{\left(  r_{BAO}-r_{BAO}^{obs}\right)  ^{2}}%
{\sigma_{r_{BAO}}^{2}}\text{ .}%
\]

Finally, we add the 397 supernovae data from Constitution compilation
\cite{constitution}. Defining the distance modulus
\[
\mu(z)=5log_{10}\left[  c(1+z)\int_{0}^{z}\frac{dz^{\prime}}{E(z^{\prime}%
)}\right]  +25-5log_{10}H_{0}\text{ ,}%
\]
we have the contribution
\[
\chi_{SN}^{2}=\sum_{j=1}^{397}\frac{[\mu(z_{j})-\mu_{obs}(z_{j})]^{2}}%
{\sigma_{j}^{2}}\text{ .}%
\]
Using the expression $\chi^{2}=\chi_{lbt}^{2}+\chi_{CMB}^{2}+\chi_{BAO}%
^{2}+\chi_{r_{BAO}}^{2}+\chi_{SN}^{2}$, the likelihood function is given by%
\[
\mathcal{L}(\delta,c,h,\Omega_{\phi_{0}})\propto exp[-\frac{\chi^{2}%
(\delta,c,h,\Omega_{\phi_{0}})}{2}]\ \text{.}%
\]

In table 1 we present the values of the individual best fit parameters, with
respective $1\sigma$, $2\sigma$ and $3\sigma$ confidence intervals.

\begin{center}
\textbf{Table 1}: Values of the model parameters from lookback time, BAO, CMB
and SNe Ia.\newline\newline%
\begin{tabular}
[c]{|c|c|}\hline
$\frac{\delta}{H_{0}}$ & $-0.066\pm0.092\pm0.185\pm0.277$\\\hline
$c$ & $0.865_{-0.022-0.035-0.045}^{+0.056+0.141+0.260}$\\\hline
$\Omega_{\phi0}$ & $0.720\pm0.016\pm0.033\pm0.049$\\\hline
$h$ & $0.666\pm0.013\pm0.026\pm0.038$\\\hline
\end{tabular}

\end{center}

\bigskip

\begin{figure}[pth]
\begin{center}
\includegraphics[width=7cm,height=6cm]{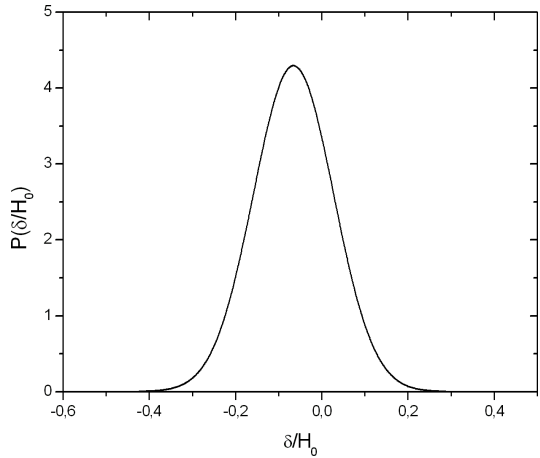}
\end{center}
\caption{Probability distribuction of $\delta$.}%
\label{pdelta}%
\end{figure}

\begin{figure}[pth]
\begin{center}
\includegraphics[width=7cm,height=6cm]{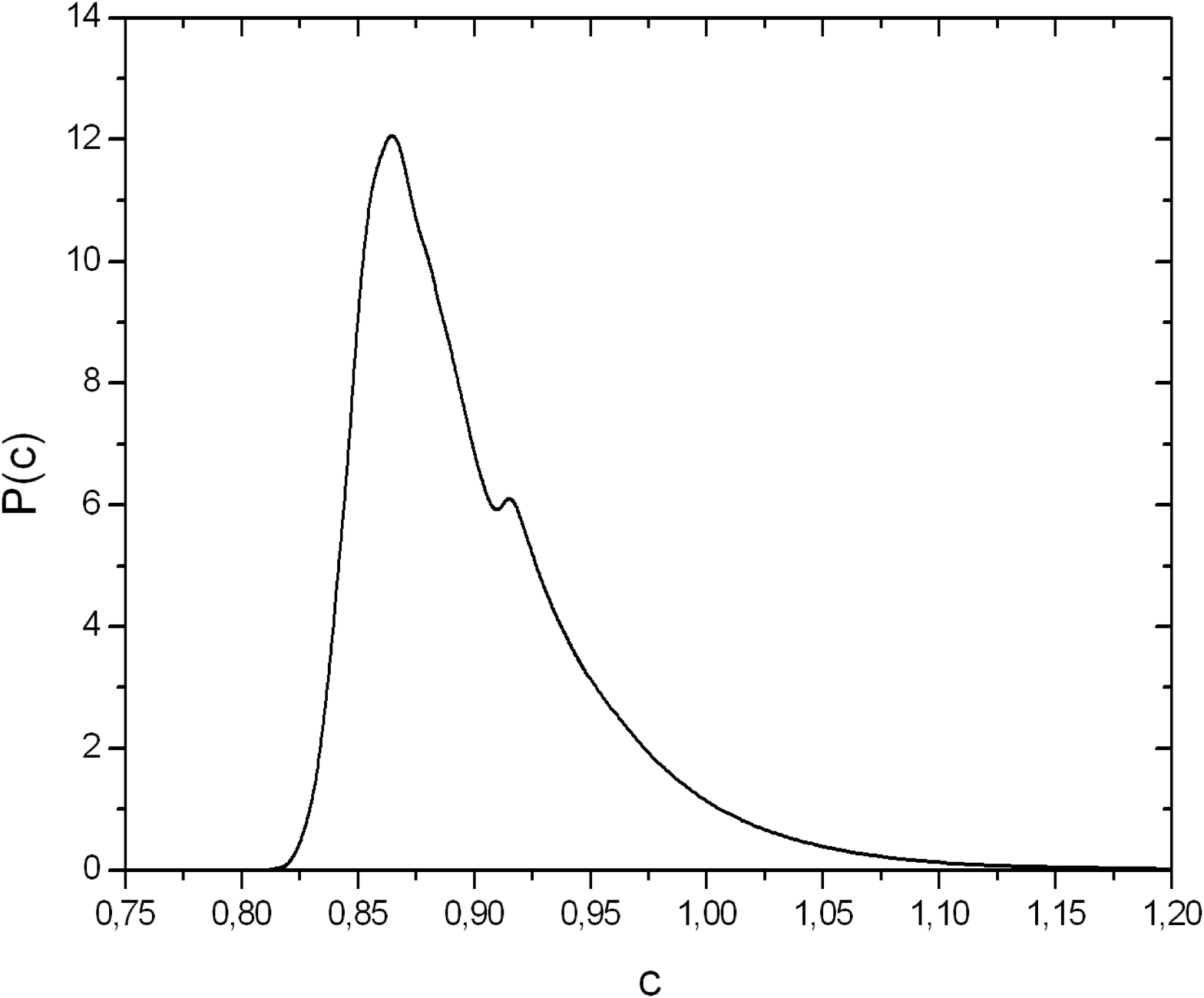}
\end{center}
\caption{Probability distribuction of $c$.}%
\label{pc}%
\end{figure}

The figures \ref{pdelta} and \ref{pc} show the marginalized probability
distributions for $\delta$ and $c$. The coupling constant $\delta$ is
compatible with zero within $1\sigma$ level. Therefore, in this work it was
not found evidence for interaction. However, the best fit value is of same
order of magnitude obtained in a previous work, where the interacting tachyon
model with power law potential has been considered \cite{sandro2}.
$\Omega_{\phi0}$ corresponds to a density matter parameter today $\Omega
_{m0}=0.280\pm0.016$, in perfect agreement with cosmological model independent
estimatives, as for example $\Omega_{Mobs}=0.28\pm0.04$ \cite{riess}. The
value of $h=0.666\pm0.013$ is also in excellent agreement with observational
values, independent of cosmological model ($h_{obs}=0.69\pm0.12$ \cite{age35}
and $h_{obs}=0.72\pm0.08$ \cite{key}).\newline

\begin{figure}[pth]
\includegraphics[width=7cm,height=6cm]{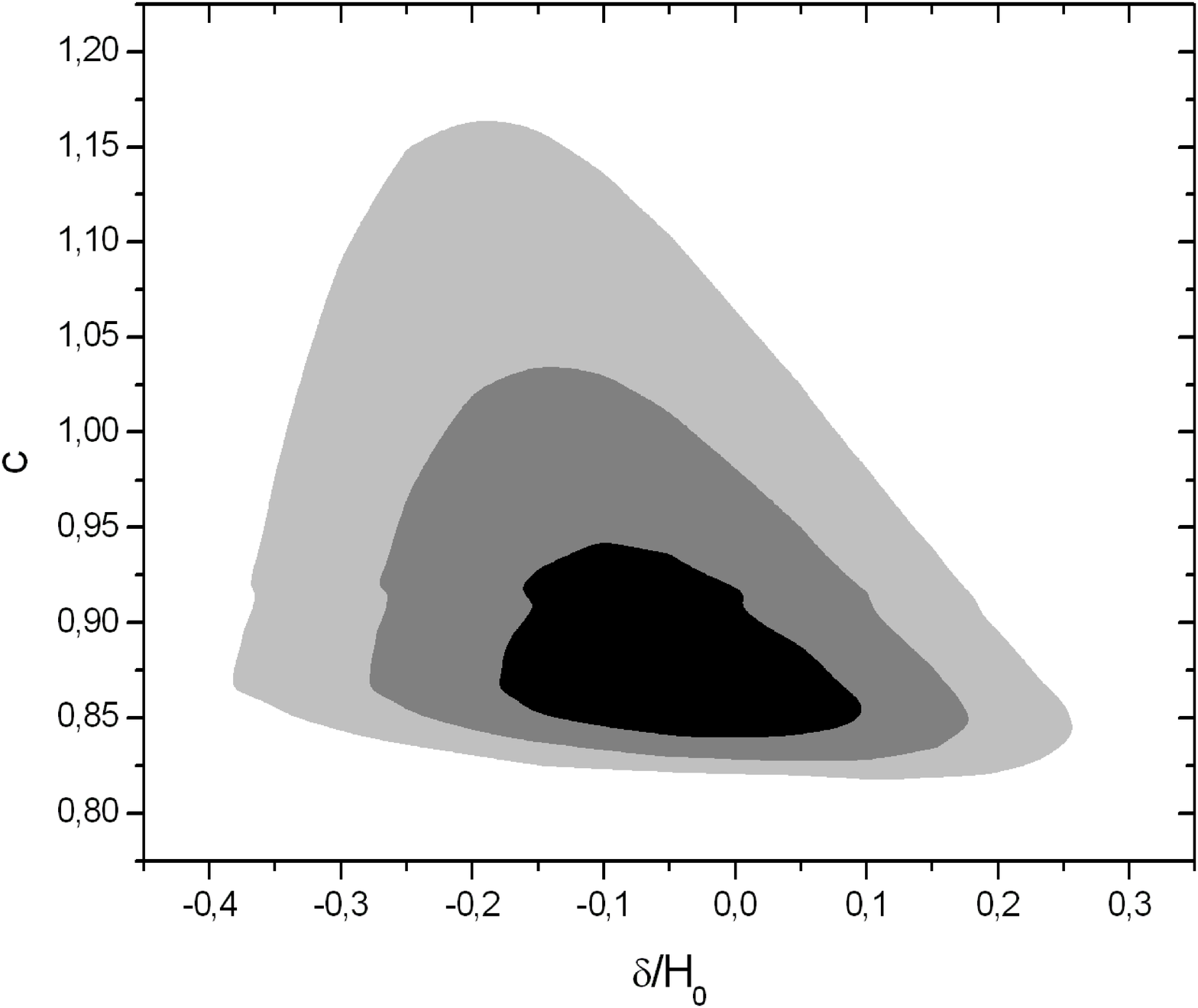}
\includegraphics[width=7cm,height=6cm]{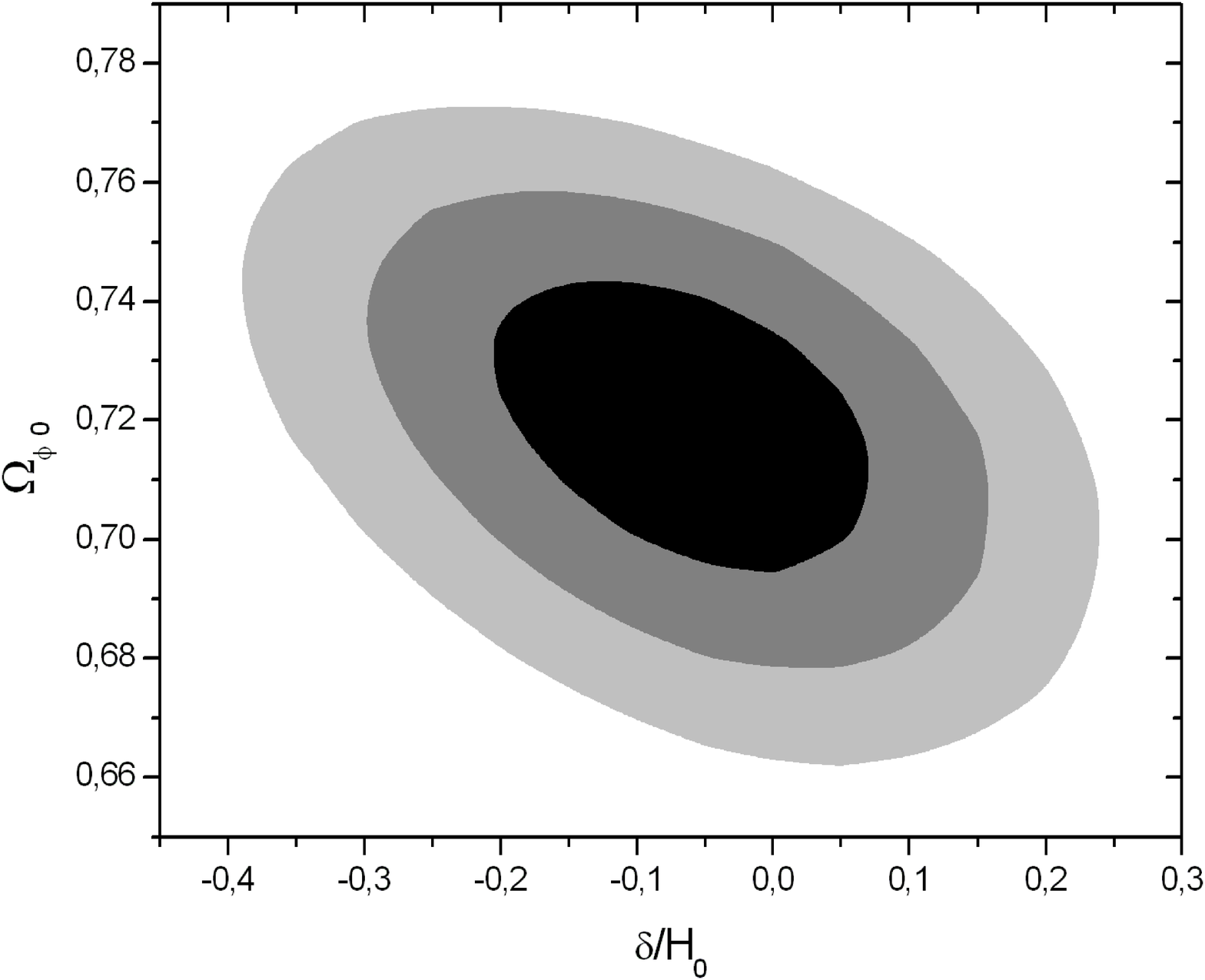}
\includegraphics[width=7cm,height=6cm]{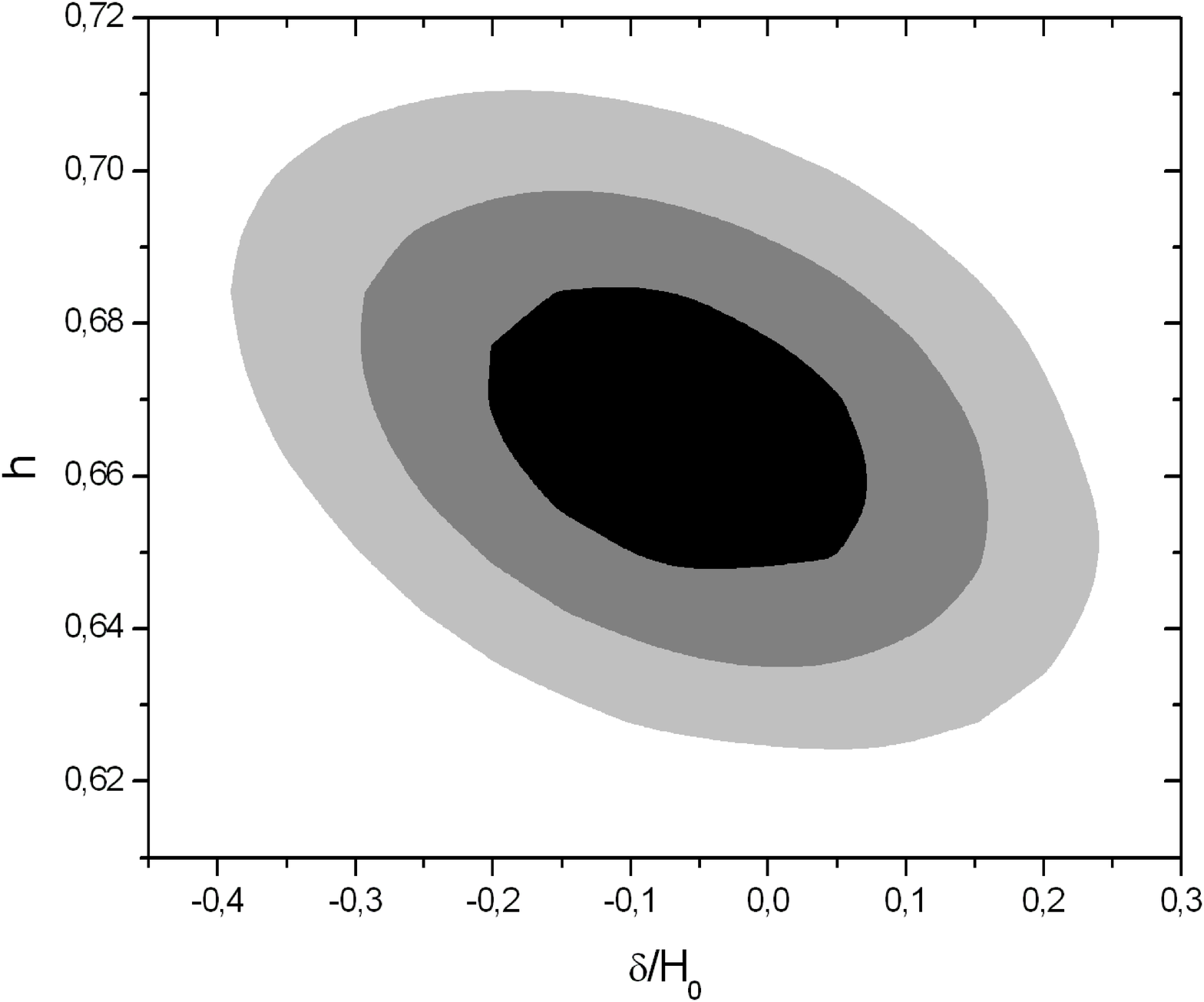}
\includegraphics[width=7cm,height=6cm]{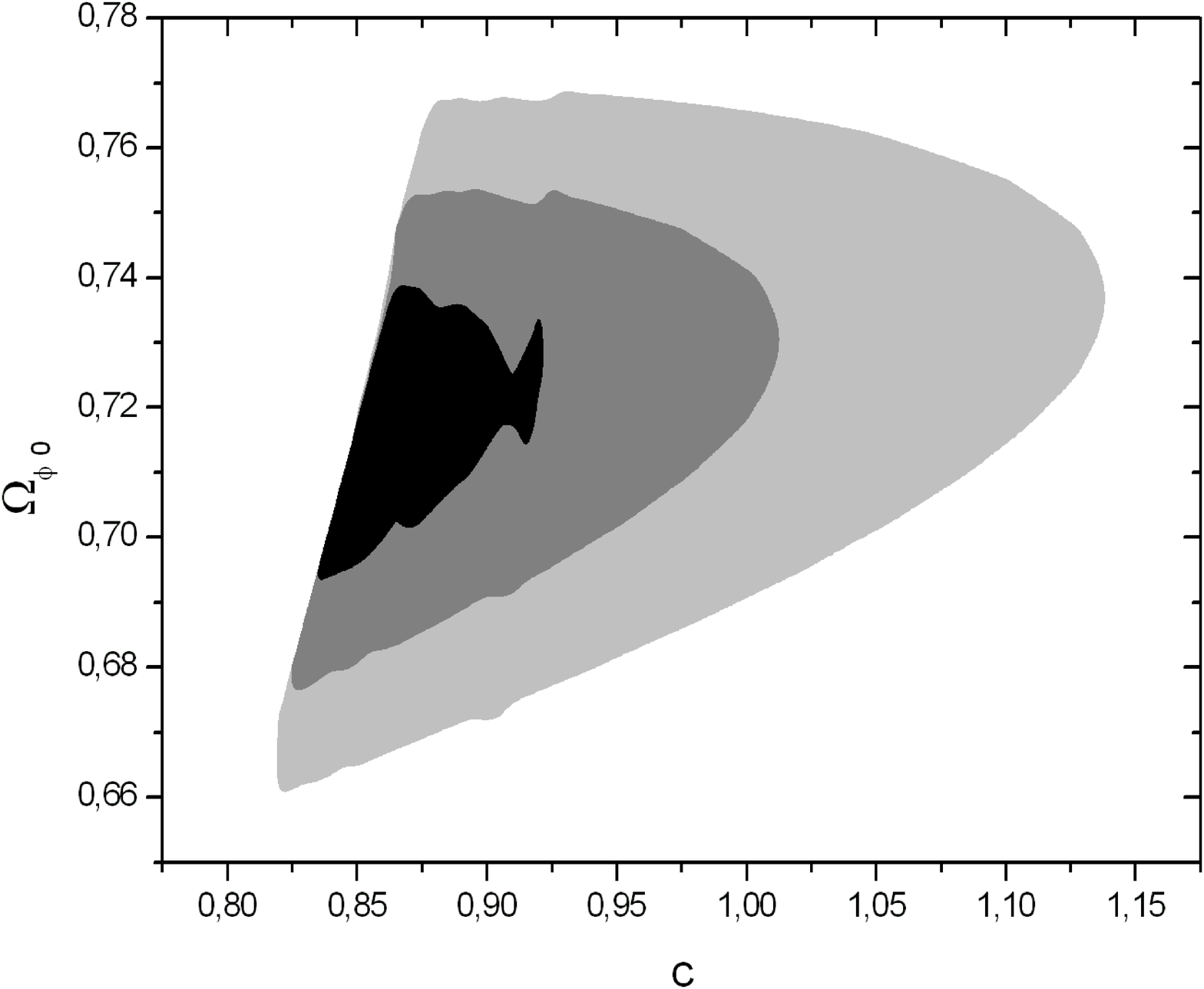} \caption{Confidence regions of
$1\sigma$, $2\sigma$ and $3\sigma$ for two parameters.}%
\label{confidence}%
\end{figure}

\begin{figure}[pth]
\begin{center}
\includegraphics[width=8cm,height=6.5cm]{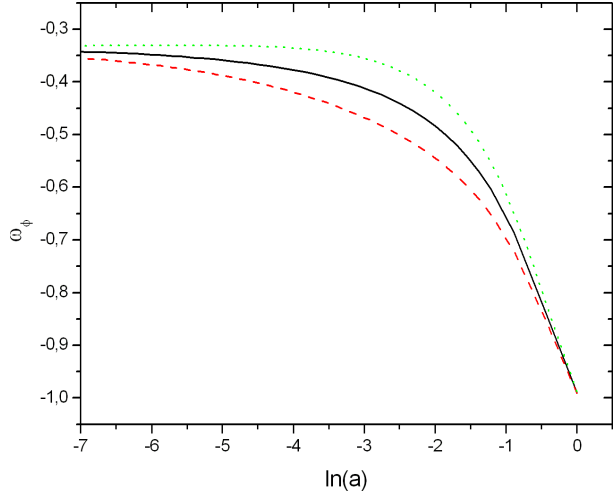}
\end{center}
\caption{Equation of state parameter of dark energy $\omega_{\phi}(a)$, for
$c=0.85$ and $\frac{\delta}{H_{0}}=-0.1$ (red dashed line), $\frac{\delta
}{H_{0}}=0$ (black solid line) and $\frac{\delta}{H_{0}}=+0.1$ (green dotted
line).}%
\label{fig_eqestado}%
\end{figure}

Figure \ref{confidence} shows the joint confidence regions for two parameters.
As we can see, there is little degeneracy between the parameters of the model.
In the confidence regions for $\delta$ versus $c$ and for $c$ versus
$\Omega_{\phi0}$, we see that there is a lower limit on $c$ $\approx0.8$. This
also can be seen in the marginalized probability distribution of $c$, which
dies for $c\lesssim0.85$. This lower limit is explained by the condition
$\frac{\sqrt{\Omega_{\phi0}}}{c}<1$, necessary for $\omega_{\phi}$ to be real
and $\omega_{\phi}>-1$, discussed above. This limit can be seen more clearly
in $c$ versus $\Omega_{\phi0}$ confidence regions. Moreover, we have
$c\simeq\sqrt{\Omega_{\phi0}}$ for the best fit values of these parameters.
This implies that $\omega_{\phi0}\simeq-1$ and the model approaches $\Lambda
CDM$ today. This is consistent with the fact that, as $\Lambda CDM$ fits all
observational data, then any alternative model must not deviates much from
$\Lambda CDM$ for $z\approx0$. However, for $z>0$, the model is qualitatively
different from $\Lambda CDM$, as $\omega_{\phi}$ approximates $-1/3$, see
figure \ref{fig_eqestado}.

We have obtained $c<1$ at $1\sigma$ confidence level. As already said above,
this implies that the equation of state parameter $\omega_{\phi}$ will not be
real for all future times. However, this is not a very serious problem,
because $c$ is compatible with values above unit at $2\sigma$ confidence
level. Moreover, one could say that $c<1$ is only an effect due to lack of
more precise observational data. Anyway, the very simple model presented here
is expected to be only an effective description of a more sophisticated
subjacent theory of dark energy. In principle, nothing guarantees that it will
be a good description for all future times.\newline

In summary, a combination of holographic dark energy model and interacting
tachyon field was implemented. It was showed that it is possible to fix the
potential of interacting tachyon by imposing that the energy density of the
tachyons must match the energy density of the holographic dark energy. A
comparison of the model with recent observational data was made and the
coupling is consistent with zero. However, in a previous work \cite{sandro2}
with interacting tachyons with power law potential, a non-vanishing coupling
constant had been obtained with $90\%$ of confidence. So the possibility of a
small, but calculable, interaction in the dark sector remains open, and future
investigations, with more realistic models and more observational data are
necessary to solve this question.

\begin{center}
\bigskip

\textbf{Acknowledgements}
\end{center}

This work has been supported by CNPq (Conselho Nacional de Desenvolvimento
Cient\'{\i}fico e Tecnol\'{o}gico) of Brazil. We would like to thank E.
Abdalla for suggestions and useful comments and Y. Gong for useful conversations.

\end{document}